\documentclass[epj,twocolumn]{webofc}
\usepackage[varg]{txfonts}   
\woctitle{NSRT15}
\newcommand{\beq}{\begin{equation}}
\newcommand{\eeq}{\end{equation}}
\newcommand{\bea}{\begin{eqnarray}}
\newcommand{\eea}{\end{eqnarray}}

\newcommand{\mkrm}[1]{}           

\begin{document}
\title{Fayans functional for deformed nuclei. Uranium region.}

\author {S.\,V. Tolokonnikov\inst{1,2} \fnsep\thanks{\email{tolkn@mail.ru}}
\and I.\,N. Borzov\inst{1,3} \and  M. Kortelainen\inst{4,5} \and Yu.\,S. Lutostansky\inst{1}
\and E.\,E. Saperstein\inst{1,6} \fnsep\thanks{\email{saper43_7@mail.ru}} }
\institute{ National Research Centre ``Kurchatov Institute'', 123182, Moscow, Russia
\and  Moscow Institute of Physics and Technology, 141700, Dolgoprudny, Moscow Region, Russia
\and   Joint Institute for Nuclear Research, 141980 Dubna, Russia
\and Department of Physics, P.O. Box 35 (YFL), University of Jyvaskyla, FI-40014 Jyvaskyla, Finland
\and Helsinki Institute of Physics, P.O. Box 64, FI-00014 University of Helsinki, Finland
\and National Research Nuclear University MEPhI, 115409 Moscow, Russia }

\abstract {Fayans energy density functional (EDF) FaNDF$^0$  has been  applied to  the  nuclei  around
uranium region. Ground state characteristics  of the Th, U and Pu isotopic chains,  up to the
two-neutron drip line, are found and compared with predictions from several Skyrme
EDFs. The two-neutron drip line is found for FaNDF$^0$, SLy4 and SkM* EDFs for a set of elements with
even proton number, from Pb  up to  Fm. }

\maketitle

\section{Introduction}
 Presently,  the Hartree-Fock (HF)  or Hartree-Fock-Bogo\-liubov (HFB) methods
together with the effective Skyrme forces \cite{HF-VB}, Gogny forces \cite{Gogny}
 or relativistic mean-field (RMF) models \cite{RMF} are most common
microscopical models applied for description of the  ground state properties of  the  heavy nuclei.
 All of the aforementioned
approaches are usually interpreted as a version of the energy density functional (EDF) method
suggested by Kohn and Sham \cite{K-Sh}. This method is based on the theorem of Hohenberg - Kohn
\cite{Hoh-K}, which states that the ground state energy $E_0$ of any quantum system is a functional of
its density $\rho({\bf r})$. By itself, the theorem says nothing about the form of this functional,
and various options for the Skyrme EDF and the Gogny EDF are, in fact, different ``ansatzs''. Among
popular Skyrme EDFs, there are quite old functionals SKM* \cite{skms} and SLy4 \cite{sly4to7}, see
the review article \cite{Ben03}. A record in the accuracy of the description of the nuclear masses
belongs to the Skyrme functional HFB-17 \cite{HFB-17} with the  average deviation from the  experiment
being  around  600 keV,  obtained by adding phenomenological corrections atop of the mean-field.
Of a comparable accuracy are the other functionals of this family, up to the
HFB-27, the corresponding tables   are presented  on the site \cite{site}.   Also relatively new
functional UNEDF1 \cite{UNE1}  has proved to be  very successful in describing the  deformed nuclei.

It is  also  worth to mention a relatively newly developed approach, known originally as a
BCP (Barcelona - Catania - Paris) \cite{BCPM1}  method, and  later  as a
BCPM (Barcelona - Catania - Paris - Madrid)  method  \cite{BCPM6}. The  main  bulk term of the BCPM functional
 was  found  by  starting from the equation of  the  state of nuclear and neutron matter,  obtained  within the
Brueckner--Hartree--Fock method  by  using a realistic  $NN+NNN$  potential.

We use the  EDF  developed by S.A. Fayans with coauthors \cite{Fay1,Fay4,Fay5,Fay}. In comparison to
the Skyrme or Gogny  EDFs, it possess two main peculiarities. Firstly, the main in-volume term of
the Fayans EDF has more sophisticated form. It can be  schematically  written as \beq{ \cal E}(\rho) =
\frac{a \rho^2} 2 \frac{1+ \alpha \rho^{\sigma}}{1+ \gamma \rho^{\sigma}}.\eeq The corresponding term
of the Skyrme EDF  would  correspond to $\gamma{=}0$  case within  this relation. The use of the bare mass, $m^*{=}m$, is
another peculiarity of the Fayans method. Both features of this approach are closely related to
the self-consistent Theory of Finite Fermi Systems (TFFS) \cite{Khod-Sap}. The latter is based on
general principles of the TFFS \cite{AB1} supplemented with the TFFS self-consistency relations
\cite{Fay-Khod}. These two peculiarities of the Fayans approach reflect, in a  effective manner, energy
dependence effects inherent to the self-consistent TFFS. E.g., the effective mass in this approach is
a product $m^*{=}m^*_k \cdot m^*_E$ of the ``$k$-mass'' and the ``$E$-mass''. The two effects
compensate each other almost exactly \cite{Khod-Sap} resulting in $m^* \simeq 1$. In the Skyrme HF
method, the $E$-mass is identically equal to unit and the effective mass may  deviate from  unity.

 Until recently, the Fayans method was  applied  for spherical nuclei only. It turned out  to be  rather successful
in systematic description of nuclear magnetic \cite{mu1,mu2}  and quad\-ru\-pole \cite{QEPJ,QEPJ-Web} moments,
nuclear radii \cite{Sap-Tolk}, beta-decay probabilities \cite{Borz},
the energies and $B(E2)$ values for the first  excited  $2^+$  states  in semi-magic nuclei
\cite{BE2,BE2-Web}. In a recent study of the single-particle energies in seven  magic nuclei, a record
accuracy was achieved \cite{Levels}. Most of these calculations  were  made  with  the EDF DF3-a  \cite{Tol-Sap}, which is a
small modification of the DF3 \cite{Fay4,Fay}  parameter set,  concerning the spin-dependent terms of the EDF. In particular, the effective
tensor term of DF3-a is significantly stronger than the DF3 one.

The Fayans EDF was applied recently  to deformed nuclei for the first time
\cite{Fay-def}. A localized version FaNDF$^0$ \cite{Fay5} of the general finite range Fayans EDF was used
which makes its surface term  more similar  to the Skyrme one. This allowed to
employ the computer code HFBTHO~\cite{code}, developed for Skyrme EDFs, with some modifications.

The results obtained with the FaNDF$^0$  EDF
for Pb and U  isotopic chains  turned
out to be promising.  In the present work, we continue these calculations
 around the uranium region. In particular, we  investigate  the  two-neutron drip line for a set of elements from Pb
 up to  Fm, those with even  proton number  being considered. For a systematic comparison, we carried out
also all calculations for two popular Skyrme EDFs, SkM* \cite{skms} and SLy4 \cite{sly4to7}. For
completeness, we included also into the  analysis predictions from the HFB-17 and
HFB-27 functionals taken from  Ref.~\cite{site}.  In addition we discuss deformation characteristics
of the drip  line nuclei.

\begin{figure}[b]
\resizebox{0.82\columnwidth}{!} {\includegraphics {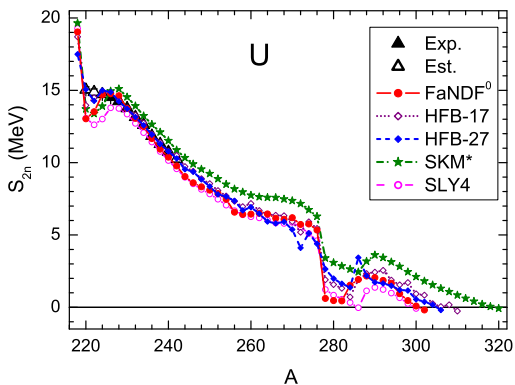}} \caption{$S_{2n}$ values in
the uranium chain for different EDFs. Experimental and estimated values are taken from \cite{mass}.} \label{fig:S2n_U}
\end{figure}
\vskip -0mm

Recent interest to the problem of fixing the neutron drip line \cite{Erl12,RMF-drip} is partially
induced with importance of this characteristic of  the nuclear chart for analysis of the r-process
dynamics in stars \cite{Kajino,Pan1,Pan2}. A  couple of remarks  should be made concerning validity of
the EDF method with fixed set of parameters for predicting the drip lines.  In particularly for the
older generation EDFs, the parameterizations were obtained by using data on nuclei close to the
stability, that is,  with the chemical potential $\mu_{n,p}{\simeq} -8\;$MeV.  When extrapolating to
very neutron rich systems, the role of EDF parameter errors becomes prominent \cite{Erl12,Kor15}.
Also,  analysis of the problem on the  basis  of the many-body theory  point of view
\cite{mu-dep1,mu-dep2} shows that in vicinity of drip lines,  that is  $\mu_n \to 0$ or $\mu_p \to 0$,
the EDF parameters describing the effective $NN$-interaction at the nuclear surface  may need to be
modified.   Close to the neutron drip line, with a small chemical potential $\mu_n$, attractive part
of the $NN$-interaction may become enhanced, resulting a deeper neutron mean field. This, as a
consequence, shifts the drip line farther away.  To take this effect into account, a simple model was
used in \cite{mu-dep1,mu-dep2} which, evidently, overestimates the effect, predicting unrealistically
strong shift of the drip line.  It was, nevertheless, demonstrated that this effect shifts the drip
line. However, one should bear in mind that by explicitly introducing such kind of component to the
EDF model, the model parameters need to be readjusted, which, so far, has not yet been done. Thus, the
total net effect remains still unknown.  In  the present  work we use the standard approach with fixed
EDF parameters.

\section{Deformation properties of Th, U, and Pu isotopic chains}
Details of the computation scheme are  identical to those described in \cite{Fay-def}. We employ the axial computer code
with the oscillator basis, the number of the oscillator shells being equal to $N_{\rm sh}{=}25$. We
limit ourselves to the quadrupole deformation $\beta_2$ only, with reflection symmetry assumed. All
the parameters of the normal component of the  used  FaNDF$^0$  EDF  are the same as in \cite{Fay5}. As  for
the anomalous term of the EDF \cite{Fay5},  \beq {\cal E}_{\rm anom}=  \sum_{i=n,p} \nu_i^{\dag}({\bf
r}) F^{\xi}(\rho_+({\bf r}))\nu_i({\bf r}), \label{Eanom}\eeq where $\nu_i({\bf r})$ is the anomalous
density, a simplified version is used, \beq F^{\xi}(\rho_+)= C_0 \left( f^{\xi}_{\rm ex} + h^{\xi}
(\rho_+/\rho_0) \right).   \label{fksi}\eeq Here $\rho_+{=}\rho_n{+}\rho_p$, and $C_0{=}\pi^2/p_{\rm
F}m$ is the usual for TFFS normalization factor. The  HFB  equations  are  solved  by using quasi-particle
cut-off energy $E_{\rm cut}\,=\,60$\,MeV.  The main part of calculations  are  carried out  by employing
simplest (``volume'') model of pairing, $h^{\xi}\,=\,0$, with $f^{\xi}_{\rm ex}\,=\,-0.440$.
 All calculations are repeated for two Skyrme EDFs, SKM* and SLy4, the results being compared also with
predictions \cite{site} from another two Skyrme EDFs,  HFB-17 and HFB-27 EDFs.

\begin{figure}
\resizebox{0.80\columnwidth}{!} {\includegraphics {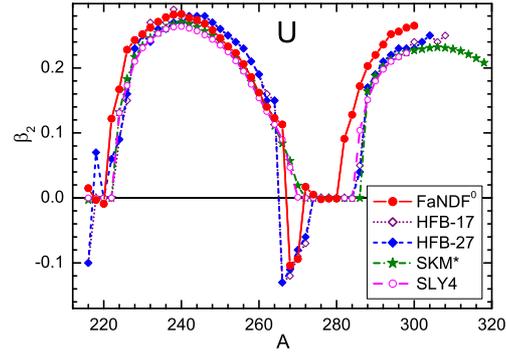}} \caption{$\beta_2$ values
in the uranium chain for different EDFs.} \label{fig:bet2_U}
\end{figure}
\vskip -0mm

Let us begin from the uranium chain. The two-neutron separation energies, \beq S_{2n}(N,Z) = B(N,Z) -
B(N-2,Z), \label{S2n} \eeq are displayed in Fig. \ref{fig:S2n_U}.  Here $B(N,Z)$  is the binding
energy of the nucleus under consideration. Comparison is made with experimental data
\cite{mass} and predictions from four Skyrme EDFs.  We first  consider region of $A
\leq 244$ with known experimental values. The HFB-17 and HFB-27  models  reaffirm their high accuracy.
As to the FaNDF$^0$ functional, agreement also looks rather reasonable, taking into account that the
parameters were fitted \cite{Fay5}  only for spherical nuclei  not heavier than lead. The deviation of
0.5\,MeV from the experimental $S_{2n}$ values for heavy U isotopes is explained mainly by  two
reasons, with is the use of a simple volume pairing interaction, and absence of the effective tensor
term in the FaNDF$^0$ EDF. Indeed, as it was shown in \cite{Tol-Sap}, the tensor term is especially
important in uranium and transuranium region as, in  corresponding  spherical case, high-$j$ levels
dominate in vicinity of the Fermi level for these nuclei. As a result, the spin-orbit density, which
comes to the EDF together with the tensor force, is typically large in these nuclei, changing
significantly along the isotopic chain. For the SLy4 EDF agreement is a bit worse, whereas the
disagreement is  more  significant in the SKM* case. It is worth to note that this Skyrme EDF, being
fitted to masses not with so high accuracy as some modern equivalents,  reproduces e.g. fission
properties in actinides or  energies and $B(E2)$ values of the first  excited  $2^+$  states in
semi-magic nuclei \cite{BE2-HFB}
 relatively well. For higher $A$ values the SKM* values of $S_{2n}$ are significantly higher
than those for all other EDFs. As a result the corresponding  drip line point  $A_{2n}^{\rm
drip}{\simeq} 320$ turns out  be  significantly  farther away  than for all other EDFs. This quantity
is defined as the last  nucleus  for which the two-neutron separation energy is yet positive.
 For SLy4 we note that $S_{2n}$ at $A=286$ is slightly negative. However, after this point the
$S_{2n}$ remain positive up to $A=298$. This kind binding re-entrance was also predicted at
\cite{Erl12}. Thus, we can set $A_{2n}^{\rm drip}({\rm SLy}4) = 298$.

\begin{figure}
\resizebox{0.80\columnwidth}{!} {\includegraphics[] {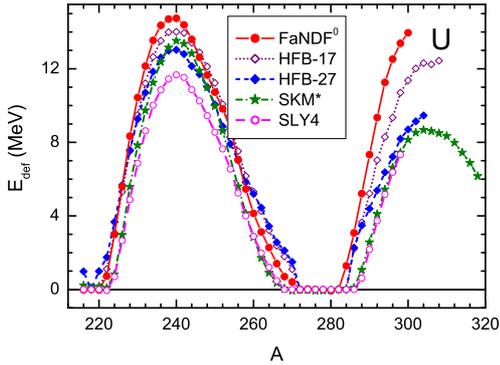}} \caption{$E_{\rm def}$
values in the uranium chain for different EDFs.} \label{fig:Edef_U}
\end{figure}
\vskip -0mm

The ground state quadrupole deformation parameter $\beta_2$ of the U chain and the corresponding
deformation energy, \beq E_{\rm def}(\beta_2) = B(\beta_2) - B(\beta_2=0), \label{Edef} \eeq are
displayed in Figs. \ref{fig:bet2_U} and \ref{fig:Edef_U}, correspondingly.
 Here, each
curve is cut around the corresponding drip point.  Generally, all five EDFs under discussion behave in
similar way, the most of nuclei in this chain being deformed in a prolate way whereas there
 exists  a region of spherical nuclei around $A{\simeq} 280$. In more detail, the width of the
spherical region is  narrowest  for the FaNDF$^0$ EDF and  widest  for SLy4. Both \ of the  HFB EDFs
predict oblate deformations for four nuclei $262{\leq} A {\leq} 268$, whereas the FaNDF$^0$ EDF, only
for two of them, $A{=}264$ and $A{=}266$.

The reason  for this  difference is quite simple. In vicinity of the phase transition, with  a  change
of the deformation sign, there are typically two  energetically close  by  minima,  the  prolate and
oblate  one. Their closeness may be confirmed with the observation that the deformation energy,
see Fig. \ref{fig:Edef_U}, does not show any non-regular behavior in the transition point,
where the  order of two minima changes. In such  kind of  situation, the transition value of $A$ may
move even due to a  small  change of the EDF parameters. In the mass region of $A > 280$ the prolate
deformation arises for the FaNDF$^0$ EDF for three points earlier than for both HFB and SLy4 EDFs. In
this region the maximum value of $\beta_2 {\simeq} 0.25$ appears for FaNDF$^0$ EDF just in the drip
 line  point. The corresponding value is a bit less for the HFB-17 EDF,  being  $\beta_2
{\simeq} 0.2$ for other three EDFs. The difference is greater for the deformation energy  due to its
quadratic behavior around the minima.  Thus, all five EDFs under consideration predict a  well
developed prolate deformation for uranium isotopes in  the vicinity of the drip line.

\begin{figure}
\resizebox{0.80\columnwidth}{!} {\includegraphics {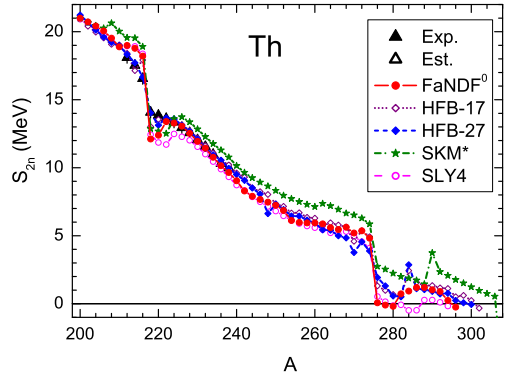}}
\vskip -1.5 mm
\caption{$S_{2n}$ values in
the thorium chain for different EDFs.} \label{fig:S2n_Th}
\end{figure}
\vskip -0mm

 Next, we investigate the thorium chain. The two-neutron separation energies are displayed  in Fig.
\ref{fig:S2n_Th}. Again, all EDFs, with exception of  SKM*, behave in general similarly in the region
of $A \preceq 280$, the latter curve being significantly higher.   Close to the drip line, however,
the picture is different. In accordance with the above discussion,  binding re-entrance can be now
seen also for FaNDF$^0$. As a  result, we obtain $A_{2n}^{\rm drip}({\rm FaNDF}^0){=}294$. For the
HFB-17 and HFB-27 EDFs, the drip points are ${\simeq} 300$ whereas again the highest value $A_{\rm
drip}{=}306$ there is for the SKM* EDF.

\begin{figure}
\resizebox{0.80\columnwidth}{!} {\includegraphics {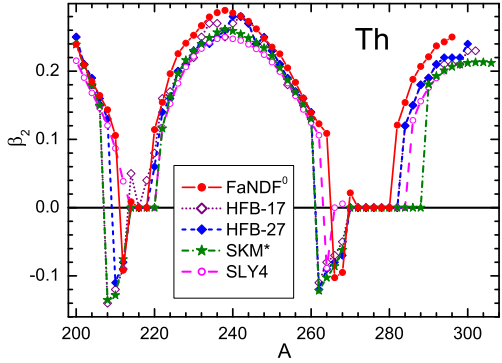}} \caption{$\beta_2$ values
in the thorium chain for different EDFs.} \label{fig:bet2_Th}
\end{figure}
\vskip -0mm

\begin{figure}
\resizebox{0.80\columnwidth}{!} {\includegraphics {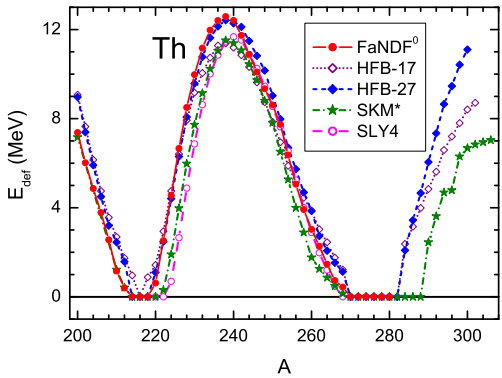}} \caption{$E_{\rm def}$
values in the thorium chain for different EDFs.} \label{fig:Edef_Th}
\end{figure}
\vskip -0mm

The ground state  deformation parameter $\beta_2$ of the Th chain and the corresponding deformation
energy are displayed in Figs. \ref{fig:bet2_Th} and \ref{fig:Edef_Th}, correspondingly. Again all five
EDFs lead to similar results  until  $A \simeq 280$, and there is rather extended region of spherical
nuclei around this point. For all EDFs under consideration, the positive deformations appear in   the
drip region, the values of deformation  parameter reaching $\beta_2\simeq 0.2$.

\begin{figure}
\resizebox{0.80\columnwidth}{!}{\includegraphics {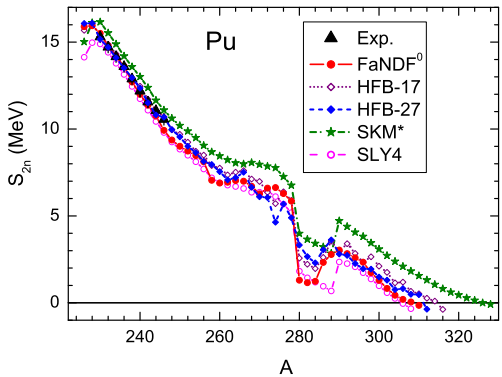}}\vskip -1mm \caption{$S_{2n}$ values in
the plutonium chain for different EDFs.} \label{fig:S2n_Pu}
\end{figure}

 Lastly, we investigate the plutonium isotopic chain.
The two-neutron separation energies are  shown  in Fig.~\ref{fig:S2n_Pu}.  Qualitatively, the picture
reminds very much that for the U case, see Fig.~\ref{fig:S2n_U}. Again, all EDFs except SKM*
reproduce reasonably the experimental data, and  for  the SKM*,  the two-neutron separation energies
are higher, especially  in  the drip region. Again  the corresponding drip  line point value $A_{\rm
drip}({\rm SKM}^*){=}324$ is much higher that those for other EDFs.

\vskip -3mm

\begin{figure}
\resizebox{0.80\columnwidth}{!} {\includegraphics[] {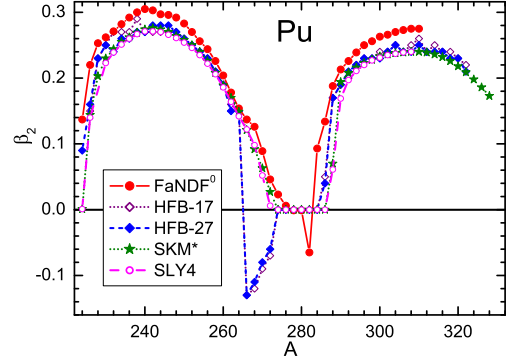}} \caption{$\beta_2$ values
in the plutonium chain for different EDFs.} \label{fig:bet2_Pu}
\end{figure}

\begin{figure}
\resizebox{0.80\columnwidth}{!} {\includegraphics {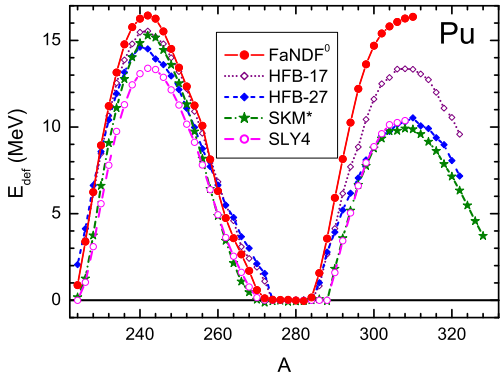}}\vskip -0mm \caption{$E_{\rm def}$
values in the plutonium chain for different EDFs.} \label{fig:Edef_Pu}
\end{figure}

The  deformation parameter $\beta_2$   and the corresponding deformation energy $E_{\rm def}(\beta_2)$
are  shown  in Figs. \ref{fig:bet2_Pu} and \ref{fig:Edef_Pu}, correspondingly. The HFB-17 and HFB-27
EDF reveal a region of oblate deformations at $266 \leq A \leq 268$. None of other EDFs confirm it.
Again, this disagreement is a consequence of a competition of two close  by  energy minima, as was
discussed  earlier. For  HFB-17 and HFB-27 EDFs, the oblate  minimum has lower energy, whereas for
other EDFs, the prolate minimum has a lower energy. Similarly as  in the uranium chain, the spherical
region ends for the FaNDF$^0$ EDF a bit earlier than for others. One more peculiarity occurs for the
Fayans EDF: after the last spherical nucleus $^{280}$Pu, a small oblate deformation $\beta_2{\simeq}
-0.05$ appears in $^{282}$Pu, which than becomes prolate  deformed, $\beta_2{\simeq} 0.1$ in
$^{284}$Pu. In this case, there is a competition of three close  by  minima, a spherical  one  and two
deformed  ones. The spherical minimum  is lowest  in $^{280}$Pu, the oblate one in $^{282}$Pu and the
prolate minimum in $^{284}$Pu. Note that  such a  small negative deformation in $^{282}$Pu is almost
 invisible  in the deformation energy  curve, in Fig. \ref{fig:Edef_Pu}.  Similarly  as in the
uranium chain, all the EDFs under consideration lead to  a strong prolate deformations in the drip
 line  region, and again the FaNDF$^0$ deformation is the  largest  one.

\vskip -0mm

\begin{table*}[ht!]
\begin{center}
\caption{ Two-neutron drip  line  points $A_{2n}^{\rm drip}(Z)$, from Pb to  Fm, for different EDFs.
 In addition, the deformation $\beta_2$ of the drip line nucleus is given in brackets.  }

\begin{tabular}{cccccc }
\hline \hline element  &FaNDF$^0$  &  SLy4 & SkM*& HFB-17 & HFB-27 \\

\hline
 Pb & 266 [0.00] &266 [0.00] & 272 [0.00] & 266 [0.00] & 266 [0.00] \\
 Po & 270 [0.00] &272 [0.00] & 280 [0.00] & 268 [0.00] & 268 [0.00] \\
 Rn & 268 [0.00] &274 [0.00] & 298 [0.20] & 272 [0.00] & 274 [0.00] \\
 Ra & 272 [0.00] &276 [0.00] & 304 [0.21] & $>$ 286    & $>$ 286    \\
 Th & 294 [0.24] &292 [0.19] & 306 [0.21] & 300 [0.23] & 298 [0.22] \\
 U  & 300 [0.26] &298 [0.22] & 318 [0.21] & 308 [0.25] & 304 [0.24]  \\
 Pu & 308 [0.27] &304 [0.24] & 324 [0.20] & 314 [0.25] & 310 [0.25]  \\
 Cm & 312 [0.28] &310 [0.25] & 336 [0.14] & 320 [0.24] & 316 [0.24] \\
 Cf & 318 [0.27] &316 [0.25] & 354 [0.00] & 324 [0.24] & 322 [0.24] \\
 Fm & 324 [0.26] &322 [0.24] & 342 [0.23] & 330 [0.23] & 328 [0.23] \\

\hline \hline
\end{tabular}
\label{tab:Q_n}
\end{center}
\end{table*}

\section{Two-neutron drip line}
In this section, we analyze the  two-neutron drip line for a set of elements with  an  even $Z$ value,
from Pb  up to  Fm. These nuclei are important for analysis of the r-process in stars
\cite{Kajino,Pan1,Pan2}. As in the previous Section, we compare predictions from the
FaNDF$^0$ functional with those from four Skyrme EDFs. The results are presented in
Table 1.
 The values in the SLy4 and SKM* columns are found using the
code \cite{code}. They coincide or are very close to the corresponding results presented in
\cite{Erl12}. Small differences  can be  explained with some distinctions in the calculation details,
 for example by the number of used oscillator shells.  The values in HFB-17 and HFB-27 are taken
from \cite{site}. For each $A_{2n}^{\rm drip}$, the corresponding value  of the deformation parameter
is given in brackets. We see that the major part of the EDFs under consideration  predict a  spherical
form  for the drip line  nuclei in the region from Pb  until  Ra, whereas all of them, from Th
 and heavier, are deformed.

As a rule, predictions from the SKM* EDF are significantly higher  compared to
others. The difference is especially large for Rn and Ra elements for which this EDF predicts a strong
deformation in the drip line region, whereas these nuclei remain to be spherical for other EDFs. For
the SKM* EDF, a strong irregularity in the $A_{2n}^{\rm drip}(Z)$ dependence occurs for Cf. The
anomalously high value of $A_{2n}^{\rm drip}({\rm SKM}^*)$ is explained with the  the competition
between the spherical and prolate minima of the total energy, the spherical one  being lower,  whereas
the prolate ones are lower in all  the neighboring elements.   The predictions from
the HFB-17 and HFB-27 EDFs are, as a rule, rather close to each other. The maximal difference between
the corresponding values of $A_{2n}^{\rm drip}$  equals  to 4.  As to the FaNDF$^0$ predictions, they
turned out to be very close to those  of the SLy4 EDF. The difference between them and both HFB EDFs
is also quite moderate.

\vskip -0mm

A remark should be made concerning the relation between $A_{2n}^{\rm drip}$ values and one-neutron
drip  line  points $A_n^{\rm drip}$. Usually, due to the pairing effect, the inequality
 $A_n^{\rm drip} {\le}A_{2n}^{\rm drip}$  is valid \cite{Erl12,RMF-drip}.

Finally, we would like to stress, that in particularly such a heavy region of the nuclear chart,
extrapolation of current EDF models up to the drip line is prone to large uncertainties. Therefore,
the results presented here should be taken with a typical uncertainty bar of $\pm 10$ mass units, or
more \cite{Erl12}. An additional shift of the two-neutron drip line may occur due to the effect of
$\mu$-dependence of the EDF parameters \cite{mu-dep1,mu-dep2},  as  discussed in the Introduction.

\section{Conclusions}

Fayans energy density functional FaNDF$^0$ is applied to nuclei  around the uranium region. For Th, U,
and Pu isotopic chains, the two-neutron separation energies $S_{2n}$, the ground state quadrupole
deformation parameter $\beta_2$  and the corresponding deformation energies $E_{\rm def}(\beta_2)$ are
found and compared with predictions from several   Skyrme EDFs. Those from
the SLy4 and SkM* EDFs
 were calculated by using the code \cite{code},  whereas the HFB-17 and HFB-27 predictions are
taken from \cite{site}. For the major part of nuclei with known experimental $S_{2n}$ values, the
results obtained with the Fayans and SLy4 EDFs are rather close to those
with two HFB EDFs, the HFB-17 one being highly accurate in the overall description of nuclear
masses \cite{HFB-17}. The SKM* EDF overestimates $S_{2n}$ values leading for these three elements to
the two-neutron drip line point $A_{2n}^{\rm drip}$ values noticeably higher than \mkrm{for}
those obtained with the other EDFs under consideration. The deformation characteristics, the
deformation parameter $\beta_2$ and the corresponding deformation energy $E_{\rm def}(\beta_2)$, for
the FaNDF$^0$ EDF, are also in the overall agreement with those of SLy4 and two the HFB EDFs. The SKM*
predictions are again different, especially  around  the drip line region.

The  two-neutron drip line is found for FaNDF$^0$, SLy4 and SkM* EDFs for a set of elements with even
 proton number, from Pb  up to Fm. This part of the nuclear chart is important for the study of
the r-process in stars \cite{Kajino,Pan1,Pan2}. The consideration is made within a standard approach
with fixed EDF parameters found mainly for stable nuclei.  Within such approach, there is an overall
agreement between all EDFs under consideration except SKM*, the latter predicting sufficiently higher
$A_{2n}^{\rm drip}$ values.

To conclude, the ground state  properties of de\-formed nuclei in the uranium region, predicted from
the FaNDF$^0$ EDF, are found to be  rather  similar to those  from several popular Skyrme EDFs. The
same also holds  for the prediction of the  two-neutron drip line for even  proton number  elements
from Pb to Fm.  With such estimates, however, one should be careful since they contain a lot of
uncertainties. Nevertheless, the FaNDF$^0$ prediction for the drip line seems to be in the line with
typical Skyrme EDFs.

\section{Acknowledgment}
The work was partly supported  by the Grant NSh-932.2014.2 of the Russian Ministry for Science and
Education, by the RFBR Grants 13-02-00085-a, 13-02-12106\_ofi-m, 14-02-00107-a, 14-22-03040\_ofi-m,
the Grant by IN2P3-RFBR under Agreement No. 110291054, and Swiss National Scientific Foundation Grant
No. IZ73Z0\_152485 SCOPES.   This work was also supported (M.K.) by Academy of Finland under the
Centre of Excellence Programme 2012--2017 (Nuclear and Accelerator Based Physics Programme at JYFL)
and FIDIPRO program. Calculations are partially made on the Computer Center of NRC ``KI''.

\vskip -15mm

\vskip -1mm
\begin{thebibliography}{99}
\bibitem{HF-VB}D. Vautherin and D. M. Brink, Phys. Rev. C {\bf 5}, 626 (1972).
\bibitem{Gogny}J. Decharg\'e and D. Gogny, Phys. Rev. C {\bf 21}, 1568
(1980).
\bibitem{RMF} P. Ring, Prog. Part. Nucl. Phys. {\bf 37}, 193 (1996).

\bibitem{K-Sh} W. Kohn and L. J. Sham, Phys. Rev.  {\bf 140}, A1133 (1965).
\bibitem{Hoh-K} P. Hohenberg and W. Kohn, Phys. Rev. {\bf 136}, B864 (1964).

\bibitem{skms} J. Bartel, P. Quentin, M. Brack, C. Guet, and H. B. H{\aa}kansson, Nucl. Phys. A
{\bf 386}, 79 (1982).
\bibitem{sly4to7}  E. Chabanat, P. Bonche, P. Haensel, J. Meyer, and R. Schaeffer, Nucl. Phys.
A {\bf 635},  231 (1998).

\bibitem{Ben03} {M. Bender, P.-H. Heenen, and P.-G. Reinhard, Rev. Mod. Phys. {\bf 75}, 121
  2003)}.

\bibitem{HFB-17} S. Goriely, N. Chamel, and J. M. Pearson, Phys. Rev. Lett.
{\bf 102}, 152503 (2009).

\bibitem{site} S. Goriely, http://www-astro.ulb.ac.be/bruslib/\\ nucdata/


\bibitem{UNE1} M. Kortelainen, J. McDonnell, W. Nazarewicz, P.-G. Reinhard, J. Sarich J,
 N. Schunck,  M. V. Stoitsov, and  S. M. Wild, Phys. Rev. C {\bf 85}, 024304 (2012).


\bibitem{BCPM1}  M. Baldo, P. Schuck,  and X. Vi$\check{\rm n}$as,   Phys. Lett.   B {\bf 663}, 390 (2008).
\bibitem{BCPM6} M. Baldo, L.  M. Robledo, P. Schuck,  and X. Vi$\check{\rm n}$as,  Phys. Rev. C {\bf 87}, 064305 (2013).


\bibitem{Fay1}  A. V. Smirnov,  S. V. Tolokonnikov, and  S. A. Fayans,  Sov. J. Nucl.
Phys. {\bf 48},  995 (1988).

\bibitem{Fay4}  I. N. Borzov, S. A. Fayans, E. Kromer, and D. Zawischa,
Z. Phys. A {\bf 355},  117 (1996).

\bibitem{Fay5} S. A. Fayans,  JETP Lett. {\bf 68}, 169 (1998).

\bibitem{Fay} S. A. Fayans, S. V. Tolokonnikov, E. L. Trykov,  and D. Zawischa,  Nucl. Phys.  A {\bf 676},  49 (2000 ).

\bibitem{Khod-Sap} V. A. Khodel, E. E. Saperstein, Phys. Rep. {\bf 92}, 183 (1982).

\bibitem{AB1} A. B. Migdal, {\it Theory of finite Fermi systems and applications to
atomic nuclei} (Nauka, Moscow, 1965; transl. Wiley, New York, 1967).

\bibitem{Fay-Khod}  S. A. Fayans and  V. A. Khodel JETP Lett. {\bf 17}, 444 (1973).

\bibitem{mu1}  I. N. Borzov, E. E. Saperstein, and S. V. Tolokonnikov,
Phys. At. Nucl. {\bf 71}, 469 (2008).
\bibitem{mu2}  I. N. Borzov, E. E. Saperstein, S. V. Tolokonnikov, G. Neyens, and N. Severijns,
Eur. Phys. J. A {\bf 45}, 159 (2010).
\bibitem{QEPJ} S. V. Tolokonnikov, S. Kamerdzhiev, S. Krewald, E. E. Saperstein, and D. Voitenkov,
Eur. Phys. J. A {\bf 48}, 70 (2012).
\bibitem{QEPJ-Web} S. Kamerdzhiev, S. Krewald, S. Tolokonnikov, E. E. Saperstein, and
D.Voitenkov.  EPJ Web of Conferences 38, 10002 (2012).
\bibitem{Sap-Tolk} E. E. Saperstein, S. V. Tolokonnikov, Phys. At. Nucl. {\bf 74}, 1277 (2011).
\bibitem{Borz}  I. N. Borzov, Phys. Rev. C {\bf 67}, 025802 (2003); Phys. Rev. C {\bf 71}, 065801 (2005).
\bibitem{BE2} S. V. Tolokonnikov, S. Kamerdzhiev, D. Voytenkov, S. Krewald, and E. E. Saperstein,
Phys. Rev. C {\bf 84}, 064324 (2011).
\bibitem{BE2-Web}S. V. Tolokonnikov, S. Kamerdzhiev,  S. Krewald, E. E. Saperstein
and D. Voitenkov.  Eur. Phys. J. Web of Conferences 38, 04002 (2012).
\bibitem{Levels} N. V. Gnezdilov, I. N. Borzov, E. E. Saperstein, and S. V. Tolokonnikov,
Phys. Rev. C {\bf 89}, 034304 (2014).

\bibitem{Tol-Sap} S. V. Tolokonnikov and E. E. Saperstein, Phys. At. Nucl. {\bf 73}, 1684 (2010).

\bibitem{Fay-def} S. V. Tolokonnikov, I. N. Borzov, M. Kortelainen, Yu. S. Lutostansky, and E. E.
Saperstein, J. Phys. G, {\bf 42}, 075102 (2015).

\bibitem{code}  M.  V. Stoitsov, N. Schunck, M. Kortelainen, N. Michel, H. Nam, E. Olsen,
J. Sarich, and S. Wild,  Comp. Phys. Comm. {\bf 184}, 1592 (2013).

\bibitem{Erl12} J. Erler, N. Birge, M. Kortelainen, W. Nazarewicz, E. Olsen, A. M. Perhac, and M. Stoitsov,
 Nature  {\bf 486},  509  (2012).

\bibitem{RMF-drip} A. V. Afanasjev,  S. E. Agbemava,  D. Ray,  and P. Ring, Phys. Rev. C {\bf 91},
014324 (2015).

\bibitem{Kajino} T. Kajino, S. Wanajo, and G. J. Mathews, Nucl. Phys. A, {\bf 704}, 165 (2002).

\bibitem{Pan1} I. V. Panov,  Astron. Lett. {\bf 29},  163  (2003).

\bibitem{Pan2} I. Petermann, K. Langanke, G. Mart\'{\i}nez-Pinedo, I. V. Panov, P.-G. Reinhard, and
F.-K.   Thielemann,  EPJ A, {\bf 48}, 122 (2012).

\bibitem{Kor15}{M. Kortelainen, J. Phys. G {\bf 42}, 034021 (2015).}

\bibitem{mu-dep1} M. Baldo, U. Lombardo, E. E. Saperstein, and
M. V. Zverev, Phys. Lett. B {\bf 533}, 17 (2002).
\bibitem{mu-dep2}  E. E. Saperstein and S. V. Tolokonnikov,  JETP Lett. {\bf 78}, 343
(2003).


\bibitem{mass}  G. Audi, A. H. Wapstra, and  C. Thibault,  Nucl. Phys. A {\bf 729},  337 (2003).

\bibitem{BE2-HFB} J. Terasaki, J. Engel, and G. F. Bertsch, Phys. Rev. C {\bf 78}, 044311 (2008).
\end{thebibliography}
\end{document}